\documentstyle[preprint,aps,epsfig]{revtex}
\newcommand{\eq}{\begin{eqnarray}}
\newcommand{\en}{\end{eqnarray}}

\tightenlines
 
\begin{document}
 
\title{Nucleon polarizabilities in the perturbative chiral quark model} 
\author{Yubing \ Dong \footnotemark[1]\footnotemark[2],  
Amand \ Faessler \footnotemark[1], 
Thomas \ Gutsche \footnotemark[1], 
Jan \ Kuckei  \footnotemark[1], \\
Valery \ E. \ Lyubovitskij  \footnotemark[1], 
Kem \ Pumsa-ard  \footnotemark[1], 
Pengnian \ Shen  \footnotemark[1]\footnotemark[2] 
\vspace*{0.4\baselineskip}}
\address{
\footnotemark[1]
Institut f\"ur Theoretische Physik, Universit\"at T\"ubingen, \\
Auf der Morgenstelle 14, D-72076 T\"ubingen, Germany
\vspace*{0.2\baselineskip}\\
\footnotemark[2]
Institute of High Energy Physics, Beijing 100049, P. R. China 
\vspace*{0.3\baselineskip}\\}
 
\date{\today}
 
\maketitle
  
\vskip 1cm

\maketitle 

\begin{abstract}
The nucleon polarizabilities $\alpha_E$ and $\beta_M$ are studied in the 
context of the perturbative chiral quark model. We demonstrate that meson 
cloud effects are sufficient to explain the electric polarizability of 
nucleon. Contributions of excite quark states to the paramagnetic 
polarizability are dominant and cancel the diamagnetic polarizability  
arising from the chiral field. The obtained results are compared to data 
and other theoretical predictions. 
\end{abstract}
 
\vskip 1cm
  
 
\vskip 1cm


\newpage 

\section{INTRODUCTION}

The nucleon polarizabilities $\alpha_N$ and $\beta_N$ with $N=p$ or $n$ 
are two of the fundamental observables which encode the electromagnetic 
structure of the nucleon. They can be determined from the differential cross 
sections of Compton scattering off the nucleon $\gamma + N \to \gamma + N$ 
(for a review see Refs.~\cite{Petrunkin:1981me}-\cite{Schumacher:2005an}).     
The idea to use Compton scattering to investigate the internal 
structure of the nucleon was formulated more than 50 years 
ago~\cite{Sachs:1950sh,Aleksandrov:1956al}. The first detailed measurement 
of the nucleon polarizabilities was performed in 
1960~\cite{Goldanski:1960gd} by analyzing elastic $\gamma p$ scattering at 
40-70 MeV. Compton scattering off the nucleon is constrained at 
very small energies by the Low-Gell-Mann-Goldberger low-energy theorem 
derived in Refs.~\cite{Low:1954kd,Gell-Mann:1954kc}. It completely fixes 
the leading $O(1)$ and the next-to-leading $O(\omega)$ structure-independent 
contribution to the expansion of the scattering amplitude in powers of the 
photon frequency~$\omega$. The leading term is known as the Thomson term and 
is expressed in terms of electron charge and nucleon mass only. The second 
term is proportional to the square of the anomalous magnetic moment of the 
nucleon. The nucleon polarizabilities characterize the quadratic terms in 
$\omega$ in the Compton scattering amplitude~\cite{Petrunkin:1961pt}, 
of which the first 
structure-dependent terms come from Rayleigh scattering. Polarizabilities 
express the deformation of particles under the influence of quasi-static 
electric and magnetic fields. In particular, in the presence of an external 
electric $\vec{E}$ or magnetic $\vec{B}$ field the quark distribution of 
the nucleon becomes distorted, leading to the induced electric 
$\vec{d} = 4 \pi \alpha_E \vec{E}$ or magnetic 
$\vec{\mu} = 4 \pi \beta_M \vec{B}$ dipole moments, where the coefficients 
$\alpha_E$ and $\beta_M$ are the electric and magnetic polarizabilities of 
the nucleon. The interaction of the electric $\vec{d}$ and magnetic 
$\vec{\mu}$ dipole moments with the electromagnetic field leads to a 
change of the corresponding effective Hamiltonian. It describes 
the interaction of nucleons with electromagnetic fields generating 
quadratic terms in $\vec{E}$ and $\vec{B}$: 
\eq
\Delta H_{\rm eff} =  - \frac{4 \pi}{2} \, ( \alpha_E \vec{E}^{\,2}  
+ \beta_M \vec{B}^{\,2} ).   
\en 
The nucleon polarizabilities can be extracted from 
the Compton scattering cross section off the nucleon by comparison with 
\eq 
\frac{d\sigma}{d\Omega} \sim  [ 
 (\alpha_E + \beta_M) (1 + \cos\theta)^2 \, + \, 
 (\alpha_E - \beta_M) (1 - \cos\theta)^2 ] + \ldots 
\en
where $\theta$ is the scattering angle. It was proposed already in the 50s and 
later indicated in experimental and theoretical studies that the meson cloud 
effects give a significant contribution to the nucleon 
polarizabilities~\cite{Schumacher:2005an}. 
The sum $\alpha_E + \beta_M$ is constrained by the so-called Baldin 
sum rule~\cite{Baldin:1960bn} which establishes a connection to the 
integral over the unpolarized photo-absorption cross section. 
The current experimental data for nucleon polarizabilities 
are~\cite{Schumacher:2005an}: 
\begin{eqnarray*}
& &\alpha_E^p = 12.0 \pm 0.6 \, \times 10^{-4} \, {\rm fm}^3\,, \,\, 
   \beta_M^p  = 1.9  \mp 0.6 \, \times 10^{-4} \, {\rm fm}^3\,, \,\, \\
& &\alpha_E^n = 12.5 \pm 1.7 \, \times 10^{-4} \, {\rm fm}^3\,, \,\, 
   \beta_M^n  = 2.7  \mp 1.8 \, \times 10^{-4} \, {\rm fm}^3\,. 
\end{eqnarray*} 
Theoretical analyses of nucleon polarizabilities were done using different 
phenomenological approaches~\cite{Petrunkin:1964pt,Hecking:1980hg} (quark, 
soliton and Skyrme models) and methods of chiral perturbation 
theory~\cite{Bernard:1991rq,Hemmert:1996rw,Lvov:1993ex}. 

A nonrelativistic approach in second-order perturbation theory gives the 
well-known sum rules for $\alpha_E = \alpha_E^{0} + \Delta\alpha_E$ and 
$\beta_M  = \beta_M^{\rm para}  + \beta_M^{\rm dia}$, where for example 
\eq 
\alpha_E^{0} = 2 \sum\limits_{n \neq 0} 
\frac{\mid <n\mid d_z\mid 0>\mid^2}{E_n-E_0}\,, \hspace*{.5cm}  
\Delta\alpha_E = Z_N^2 \frac{e^2<r_{E}^{2}>^N}{3m_N}
\en 
are the leading contribution and the retardation correction to the electric 
polarizability. The magnetic polarizability contains the para- and 
diamagnetic contributions with 
\eq 
\beta_M^{\rm para}  = 2\sum_{n\neq 0} 
\frac{\mid <n\mid \mu_z\mid 0>\mid^2}{E_n-E_0}\,, \hspace*{.5cm}  
\beta_M^{\rm dia} = 
-e^2\sum_{i=1}^3\frac{Q_i^2}{6m_i}<0\mid\rho^2_i\mid 0>
- \frac{<0\mid\vec{d}~^2\mid 0>}{2m_N}. 
\en 
Here $\mid n>$ and $E_n$ are eigenstate and energy of an excited state, 
$\vec{d}$ is the electric dipole moment, $d_z$ and $\mu_z$ are the 
$z$-components of the electric and magnetic moments, respectively, 
$Z_N$ and $m_N$ are the charge and the mass of nucleon, $Q_i$, $m_i$, 
and $\rho_i$ are the charge, mass and internal coordinate of the constituent 
quark, respectively. The problems associated with the use of the 
nonrelativistic sum rules have been discussed 
in~\cite{Lvov:1993fp,Schumacher:2005an,Hemmert:1996rw,Lvov:1993ex,Mukhopadhyay:1993zx}.  
In particular, the proton electric polarizability is predicted to be larger 
than the neutron one, which is in conflict with the central values of the 
experiment with $\alpha_E^n \ge \alpha_E^p$~\cite{Hemmert:1996rw}. It turns 
out that the inclusion of the $\Delta(1232)$-resonance is essential in the 
calculation of $\beta_M$. It gives a large positive contribution to the 
paramagnetic moment which compensates a relatively large negative diamagnetic 
polarizability~\cite{Hemmert:1996rw,Mukhopadhyay:1993zx}. Finally, the main 
problem is that the above nonrelativistic definition of the nucleon 
polarizabilities is not identical to the physical observables $\alpha_E$ and 
$\beta_M$ because of missing relativistic 
corrections~\cite{Lvov:1993fp,Schumacher:2005an}. 
Therefore, an accurate analysis of nucleon 
polarizabilities should be done within a truly relativistic approach. 

Assuming $P-, C$- and $T$-invariance, the general amplitude for Compton 
scattering can be written in terms of six real structure dependent 
functions $A_i^N(\omega, \theta)$ with $i=1 \ldots 6$, 
where $\omega=\omega^\prime$ denotes the photon energy in the 
center-of-mass (CM) frame. The latter is specified by the following 
choice of momenta of initial $(p)$ and final $(p^\prime)$ nucleons 
and incoming $(k)$ and out-coming $(k^\prime)$ photon:  
\eq 
p = ( E, - \vec{k})\,, \; \; p^\prime = ( E, - \vec{k}^{\,\prime})\,, \; \;  
k = ( \omega, \vec{k})\,, \; \; k^\prime = ( \omega, \vec{k}^{\,\prime})\,. 
\en
For the polarization vectors of the real photon field it is convenient to 
choose the Coulomb gauge which is characterized by the conditions: 
\eq 
\epsilon = (0, \vec{\epsilon}\,)\,, \hspace*{.5cm} 
\epsilon^{\ast\prime} = (0, \vec{\epsilon}^{\,\,\ast\prime}\,)\,, 
\hspace*{.5cm} 
\vec{k} \cdot \vec{\epsilon} = 0\,, \hspace*{.5cm} 
\vec{k}^{\prime} \cdot \vec{\epsilon}^{\,\,\ast\prime} = 0\,. 
\en
The $T^N$-matrix of Compton scattering is then written 
as~\cite{Ritus:1957rt} 
\eq\label{Tinv}
T^N \, = \, \sum\limits_{i=1}^{6} \, A_i^N(\omega,\theta) \, S_i\,, 
\hspace*{.5cm} 
S_1 = \vec{\epsilon}^{\,\,\ast\prime} \cdot\vec{\epsilon}\,, \hspace*{.5cm} 
S_2 = \vec{\epsilon}^{\,\,\ast\prime} \cdot\hat k \ 
\vec{\epsilon} \cdot\hat k^\prime \, \,, 
\hspace*{.5cm} \ldots 
\en 
Here, $\epsilon$, $\hat{k}$ ($\epsilon^{\ast\prime}$, $\hat{k}^{\prime}$) are 
the polarization vector and  the direction of the incoming (outgoing) photon 
with $\hat k = \vec{k}/|\vec{k}|$. The scattering angle $\theta$ is defined as 
$\cos\theta = \hat{k} \cdot \hat{k}^{\,\prime}$. In Eq.~(\ref{Tinv}) 
only the first two terms are spin-independent and others are spin-dependent 
ones which contain the nucleon spin operator $\sigma$.  
The structure functions $A_i^N$ contain contributions from the 
pion-pole (anomalous) and from the remaining (regular) terms with
\begin{eqnarray}
 A_i^N(\omega,\theta)=A_i^N(\omega,\theta)^{\pi^0-{\rm pole}}
+A_i^N(\omega,\theta)^{\rm reg},~~~i=1...6.
\end{eqnarray}
However, the $\pi^0$-pole does not contribute to the spin-independent 
structure functions $A_1^N$ and $A_2^N$. One can perform a low-energy 
expansion of the six nucleon structure functions 
$A_i^N(\omega,\theta)^{\rm reg}$ in powers of the photon energy $\omega$. The 
explicit expressions of the expansion for the case of a proton target of mass 
$M_N$ with anomalous magnetic moment $\kappa^p$ can be found in 
Refs.~\cite{Drechsel:2002ar,Schumacher:2005an}. The explicit expressions 
of the spin-independent functions $A_1^N(\omega,\theta)^{\rm reg}$ 
and $A_2^N(\omega,\theta)^{\rm reg}$, which are expressed in terms 
of $\alpha_E$ and $\beta_M$, are~\cite{Bernard:1991rq} 
\begin{eqnarray}\label{A1_A2_reg}
A_1^N(\omega,\theta)^{\rm reg} &=& 
- \, \frac{e^2}{m_N}\, Q_N^2 \, + \, 
4\, \pi\, (\alpha_E^N \, + \,\cos\theta\, \beta_M^N) 
\, \omega^2 \nonumber\\
&+& \frac{4 \, \pi}{m_N} \, (\alpha_E^N \, + \, \beta_M^N) \, 
(1 + \cos\theta) \, \omega^3 \, + \, 
O\biggl(\omega^4, \frac{1}{m_N^3}\biggr) \,, \nonumber\\
A_2^N(\omega,\theta)^{\rm reg} &=& \frac{e^2}{m_N^2} \, Q_N^2 \, 
\omega \, - \, 4 \, \pi \, \beta_N^N \,  \omega^2 
\, - \, \frac{4 \, \pi}{m_N} \, (\alpha_E^N \, + \, \beta_M^N) \, \omega^3
 \, + \, 
O\biggl(\omega^4, \frac{1}{m_N^3}\biggr) \,, 
\end{eqnarray}
where $Q_N$ is the nucleon charge. 

In this work, a perturbative chiral quark model 
(PCQM)~\cite{Lyubovitskij:2001nm,Lyubovitskij:2000sf,Cheedket:2002ik} is 
employed to study the spin-independent polarizabilities of nucleon, $\alpha_N$ 
and $\beta_N$. In the PCQM baryons are described by three relativistic 
valence quarks confined in a static potential, which are supplemented by a 
cloud of pseudoscalar Goldstone bosons, as required by chiral symmetry. This 
model has already been successfully applied to the charge and magnetic form 
factors of baryons, sigma terms, ground state masses of baryons, the 
electromagnetic $N \to \Delta$ transition, and other baryon  
properties~\cite{Lyubovitskij:2001nm,Lyubovitskij:2000sf,Cheedket:2002ik}. 
A detailed discussion of the approach and a comparison to similar chiral quark 
models can be found in Ref.~\cite{Lyubovitskij:2001nm}.  To calculate the 
nucleon polarizabilities we apply the heavy nucleon mass limit 
$m_N \to \infty$  where only the $\omega^2$ terms survive in the expansion of 
the regular amplitudes $A_{i=1,2}^N(\omega,\theta)^{\rm reg}$. 
From an evaluation of these terms we extract the nucleon polarizabilities. 

The paper is organized as follows. In Sec. II we present an introduction 
to the PCQM. In Sec. III we calculate the nucleon polarizabilities in the 
PCQM. We discuss the numerical results and compare them to the predictions of 
other theoretical approaches. In Sec. IV we give our conclusions. 

\section{THE PERTURBATIVE CHIRAL QUARK MODEL}

The perturbative chiral quark model is based on an effective chiral 
Lagrangian describing baryons by a core of the three valence quarks, 
moving in a central Dirac field with $V_{\rm eff}(r)=S(r)+\gamma^0V(r)$, 
where $r=\mid\vec{x}\mid$. In order to respect chiral symmetry, 
a cloud of Goldstone bosons ($\pi$, $K$ and $\eta$) is included as 
small fluctuations around the three-quark core. Here we restrict to the 
SU(2) version of our approach and take into account only the pion cloud 
fluctuations. The total model Lagrangian is 
\begin{eqnarray}\label{Leff}
{\cal L}_{\rm eff}={\cal L}_{\rm inv}+{\cal L}_{\chi SB},
\end{eqnarray}
where ${\cal L}_{\rm inv}$ is the chiral-invariant Lagrangian and 
${\cal L}_{\chi SB}$ is the chiral symmetry breaking part.
Up to second order in the pion field $\hat\pi = \pi_i \tau_i$ fluctuations, 
${\cal L}_{\rm inv}$ and ${\cal L}_{\chi SB}$ are
\begin{eqnarray}\label{L_eff}
{\cal L}_{\rm inv}(x) &=&
\bar{\psi}(x) [i \not\!\partial - S(r) - \gamma^0 V(r)]\psi(x) +
\frac{1}{2} [\partial_\mu \hat \pi(x)]^2 
 - \bar{\psi}(x) S(r) i \gamma^5 \frac{\hat \pi (x)}{F} \psi(x) \\ \nonumber
&+&\bar \psi (x)\frac{\hat\pi^2(x)}{2F^2}S(r)\psi (x) \, ,
\end{eqnarray}
\begin{equation}\label{Mass}
{\cal L}_{\chi SB}(x) = -\bar\psi(x) {\cal M} \psi(x)
- \frac{B}{2} Tr [\hat \pi^2(x)  {\cal M} ]\,.
\end{equation}
Eq.~(\ref{Mass}) contains the mass contributions both for the quarks and 
pions, which explicitly break chiral symmetry. Lagrangian (\ref{L_eff}) 
contains the basic parameters: $F=88$ MeV is the pion decay constant in the 
chiral limit, ${\cal M}={\rm diag}\{ \hat m,\hat m \}$ is the mass matrix of 
current quarks (we restrict to the isospin symmetry limit $m_u=m_d=\hat m$) 
and $B=-<0|\bar u u|0>/F^2 = -<0|\bar d d|0>/F^2$ is the quark condensate 
constant. We rely on the standard picture of chiral symmetry 
breaking~\cite{Gasser:1982ap} and for the masses of pseudoscalar mesons 
we use the leading term in their chiral expansion (i.e. linear in the 
current quark mass). By construction, our 
effective chiral Lagrangian is consistent with the known low-energy theorems 
(Gell-Mann-Okubo and Gell-Mann-Oakes-Renner relations, partial conservation 
of axial current (PCAC), Feynman-Hellmann relation between pion-nucleon 
$\sigma$-term and derivative of the nucleon mass, etc.). 

We expand the quark field $\psi$ in the basis of potential
eigenstates as
\begin{eqnarray}
\psi(x) = \sum\limits_\alpha b_\alpha u_\alpha(\vec{x})
\exp(-i{\cal E}_\alpha t) + \sum\limits_\beta
d_\beta^\dagger v_\beta(\vec{x}) \exp(i{\cal E}_\beta t)\, ,
\nonumber
\end{eqnarray}
where the sets of quark $\{ u_\alpha \}$ and antiquark $\{ v_\beta \}$
wave functions in orbits $\alpha$ and $\beta$ are solutions of the
Dirac equation with the static potential $V_{\rm eff}(r)$.
The expansion coefficients $b_\alpha$ and $d_\beta^\dagger$ are the
corresponding single quark annihilation and antiquark creation
operators. In our calculation of matrix elements,  we project quark diagrams 
on the respective baryon states. The baryon states are conventionally
set up by the product of the ${\rm SU(6)}$ spin-flavor and
${\rm SU(3)_c}$ color wave functions, where the nonrelativistic single
quark spin wave function is simply replaced by the relativistic solution
$u_\alpha(\vec{x})$ of the Dirac equation
\begin{eqnarray}\label{Dirac_eq}
\left[ -i\gamma^0\vec{\gamma}\cdot\vec{\nabla} + \gamma^0 S(r) + V(r)
- {\cal E}_\alpha \right] u_\alpha(\vec{x})=0,
\end{eqnarray}
where ${\cal E}_\alpha$ is the single-quark energy of state $\alpha$. 

For the description of baryon properties, we use the effective 
potential $V_{\rm eff}(r)$ with a quadratic radial 
dependence~\cite{Lyubovitskij:2001nm,Lyubovitskij:2000sf}:
\begin{eqnarray}\label{potential} 
S(r) = M_1 + c_1 r^2, \hspace*{1cm} V(r) = M_2+ c_2 r^2
\end{eqnarray}
with the particular choice
\begin{eqnarray}
M_1 = \frac{1 \, - \, 3\rho^2}{2 \, \rho R} , \hspace*{1cm}
M_2 = {\cal E}_0 - \frac{1 \, + \, 3\rho^2}{2 \, \rho R} , \hspace*{1cm}
c_1 \equiv c_2 =  \frac{\rho}{2R^3} .
\end{eqnarray}
Here, ${\cal E}_0$ is the single-quark ground-state energy;
$R$ and $\rho$ are parameters related to the ground-state quark wave
function $u_0$:
\begin{eqnarray}\label{eigenstate}
u_0(\vec{x};i) \, = \, N_0 \, \exp\biggl[-\frac{\vec{x}^{\, 2}}{2R^2}\biggr]
\, \left(
\begin{array}{c}
1\\
i \rho \, \vec{\sigma}(i)\cdot\vec{x}/R\\
\end{array}
\right)
\, \chi_s(i) \, \chi_f(i) \, \chi_c(i)\,,
\end{eqnarray}
where $N_0=[\pi^{3/2} R^3 (1+3\rho^2/2)]^{-1/2}$ is a normalization
constant; $\chi_s$, $\chi_f$, $\chi_c$ are the spin, flavor and color
quark wave functions, respectively. The index $"i"$ stands for the $i$-th 
quark. The constant part of the scalar potential $M_1$ can be interpreted as 
the constituent mass of the quark, which is simply the displacement of the 
current quark mass due to the potential $S(r)$. The parameter $\rho$ is 
related to the axial charge $g_A$ of the nucleon calculated in zeroth-order
(or 3q-core) approximation:
\begin{eqnarray}
g_A=\frac{5}{3}\biggl(1 - \frac{2\rho^2}{1+\frac{3}{2}\rho^2}\biggr)\,.
\end{eqnarray}
Therefore, $\rho$ can be replaced by $g_A$ using the matching condition (17). 
The parameter $R$ is related to the charge radius
of the proton in the zeroth-order approximation as
\begin{eqnarray}
<r^2_E>^P_{\rm LO} = \int d^3 x \, u^\dagger_0 (\vec{x}) \,
\vec{x}^{\, 2} \, u_0(\vec{x}) \, = \, \frac{3R^2}{2} \,
\frac{1 \, + \, \frac{5}{2} \, \rho^2}{1 \, + \, \frac{3}{2} \, \rho^2}.
\end{eqnarray}
In our calculations we use the value $g_A$=1.25. Therefore, we have only 
one free parameter in our model, that is $R$ or  $<r^2_E>^p_{\rm LO}$. 
In the numerical studies, $R$ is varied in the region from 0.55 fm to 0.65 fm, 
which corresponds to a change of $<r^2_E>^P_{\rm LO}$ from 0.5 to 0.7 fm$^2$. 
Note that for the given form of the effective potential~(\ref{potential}) 
the Dirac equation~(\ref{Dirac_eq}) can be solved analytically [for 
the ground state see Eq.(\ref{eigenstate}), for excited states 
see Ref.~\cite{Cheedket:2002ik}]. In the Appendix we give details of 
the solutions to the Dirac equation for any excited state. 

The expectation value of an operator $\hat A$ is then set up as:
\begin{equation}\label{hatA}
<\hat A> = {}^B\!\!< \phi_0 |\sum^{\infty}_{n=1} \frac{i^n}{n!}
\int d^4 x_1 \ldots \int d^4 x_n T[{\cal L}_I (x_1) \ldots
{\cal L}_I (x_n) \hat A]|\phi_0>^B_c,
\end{equation}
where the state vector $|\phi_0>$ corresponds to the unperturbed
three-quark state ($3q$-core). Superscript $``B"$ in the equation indicates 
that the matrix elements have to be projected onto the respective
baryon states, whereas subscript $``c"$ refers to contributions from
connected graphs only. ${\cal L}_I (x)$ of Eq.~(\ref{hatA}) refers to 
the quark-meson interaction Lagrangian:
\begin{equation}
{\cal L}_I (x) = -\bar \psi (x) i \gamma^5 \frac{\hat\pi (x)}{F}
S(r) \psi(x)+\bar \psi (x)\frac{\hat\pi^2(x)}{2F^2}
S(r)\psi (x) \,.
\end{equation}
For the evaluation of Eq.~(\ref{hatA}) we apply Wick's theorem 
with the appropriate propagators for the quarks and pions. 

For the quark field we use a Feynman propagator for a fermion in a binding 
potential with
\begin{eqnarray}\label{quark_propagator} 
i G_\psi(x,y) &=& <\phi_0|T\{\psi(x)\bar \psi(y)\}|\phi_0>\nonumber \\
&=& \theta(x_0-y_0) \sum\limits_{\alpha} u_\alpha(\vec{x})
\bar u_\alpha(\vec{y}) e^{-i{\cal E}_\alpha (x_0-y_0)}
- \theta(y_0-x_0) \sum\limits_{\beta} v_\beta(\vec{x})
\bar v_\beta(\vec{y}) e^{i{\cal E}_\beta (x_0-y_0)}\,.
\end{eqnarray} 
In Refs.~\cite{Lyubovitskij:2001nm,Lyubovitskij:2000sf} we restricted 
the expansion of the quark propagator to its ground state with:
\begin{eqnarray}
iG_\psi(x,y) \to iG_0(x,y) \doteq u_0(\vec{x}) \, \bar u_0(\vec{y}) \,
e^{-i{\cal E}_0 (x_0-y_0)} \, \theta(x_0-y_0).
\end{eqnarray}
In Ref.~\cite{Cheedket:2002ik} we investigated in addition the impact of 
excited states in the loops on the electromagnetic properties of baryons, 
the N-$\Delta$ transition, the nucleon axial vector form factors, and 
meson-baryon sigma-terms. Here,  we also include the excited quark states 
in the quark propagator of Eq.~(\ref{quark_propagator}) to analyze their 
contribution to the nucleon polarizabilities. In our numerical calculation  
we include the following set of excited quark states: the first $p$-states 
($1p_{1/2}$ and $1p_{3/2}$ in the non-relativistic notation) and the second 
excited states ($1d_{3/2}, 1d_{5/2}$ and $2s_{1/2}$), i.e. we restrict to 
the low-lying excited states with energies smaller than the typical scale of  
$\Lambda =  1$ GeV. The justification for such  an approximation can be found 
in our previous papers~\cite{Cheedket:2002ik}. 

For the pions we adopt the free Feynman propagator with
\begin{eqnarray}
i\Delta_{ij}(x-y)=<0|T\{\pi_i(x)\pi_j(y)\}|0> = 
\delta_{ij}\int\frac{d^4k}{(2\pi)^4i} \, e^{-ik(x-y)} \, \Delta_{\pi}(k) \,, 
\end{eqnarray}
where $\Delta_{\pi}(k) = [M_\pi^2-k^2-i0^+]^{-1}$ is the pion 
propagator in the momentum space. 
Introduction of the electromagnetic field $A_\mu$ to the effective
Lagrangian  is accomplished by minimal substitution:
\begin{eqnarray}
\partial_\mu\psi \to D_\mu\psi = \partial_\mu\psi + i e Q A_\mu \psi,
\hspace*{.5cm}
\partial_\mu\pi_i \to D_\mu\pi_i = \partial_\mu\pi_i +
e \varepsilon_{3ij}  A_\mu \pi_j
\end{eqnarray}
where $Q = {\rm diag}\{2/3, -1/3\}$ is the $SU(2)$ quark charge matrix
and $\varepsilon_{ijk}$ are the totally antisymmetric structure constants of
$SU(2)$ (Levi-Civita tensor).

For the photon field $A_\mu$ we also include the usual kinetic term
\begin{eqnarray}
{\cal L}_{ph} = - \frac{1}{4} F_{\mu\nu} \, F^{\mu\nu} \hspace*{.5cm}
\mbox{with} \hspace*{.5cm} F_{\mu\nu} = 
\partial_\mu A_\nu - \partial_\nu A_\mu \,.
\end{eqnarray}
The electromagnetic interaction of the system is 
\begin{eqnarray}
{\cal L}_{em} = - e \, A_{\mu} \, \bar{\psi} \, \gamma^{\mu} \, Q \, \psi 
- e \, A_{\mu} \, \varepsilon_{3ij} \, \pi_i\partial^{\mu} \, \pi_j 
+ \frac{e^2}{2} \, A_{\mu} \, A^{\mu} \, ( \vec{\pi}^2 - \pi_0^2 ) \,. 
\end{eqnarray}

\section{NUCLEON POLARIZABILITIES IN THE PCQM}

In this Section we apply the PCQM to calculate the spin-independent 
polarizabilities $\alpha_E$ and $\beta_M$  contained in the real Compton 
scattering amplitudes $A_1^N(\omega,\theta)^{\rm reg}$ and 
$A_2^N(\omega,\theta)^{\rm reg}$. In the heavy nucleon mass limit with 
$m_N \to \infty$ the structures of 
$A_1^N(\omega,\theta)^{\rm reg}$ and $A_2^N(\omega,\theta)^{\rm reg}$ 
amplitudes are sufficiently simplified as: 
\begin{eqnarray}\label{A1_A2_reg_limit}
A_1^N(\omega,\theta)^{\rm reg}\bigg|_{m_N \to \infty} &=& 
4\, \pi\, (\alpha_E^N \, + \,\cos\theta\, \beta_M^N) 
\, \omega^2, \nonumber\\
A_2^N(\omega,\theta)^{\rm reg}\bigg|_{m_N \to \infty} &=& 
\, - \, 4 \, \pi \, \beta_N^N \,  \omega^2. 
\end{eqnarray} 
This limit it well-justified in the PCQM and corresponds to the static 
three-quark bag with projection onto a nucleon state with finite 
three-momentum. Let us stress again that we do not intend to calculate  
the full Compton scattering amplitude $\gamma + N \to \gamma + N$ 
consistent with the relativistic theory. We just restrict to the evaluation 
of the spin-independent polarizabilities $\alpha_N$ and $\beta_N$. In the 
calculation of $\alpha_N$ and $\beta_N$ we proceed as follows. First we 
calculate the quark (one- and two-body) operators describing Compton 
scattering off a quark: $\gamma + q \to \gamma + q$. Then we project the set 
of these operators onto the nucleon states and extract  $\alpha_N$ and 
$\beta_N$ from matching the result to Eq.~(\ref{A1_A2_reg_limit}). 

The Feynman diagrams contributing to the process $\gamma + q \to \gamma + q$ 
(Compton scattering on the quark level) are shown in Figs.1-8:  tree-level 
diagrams (Fig.1), one-body meson-cloud diagrams (Figs.2, 3, 4, 5 and 6) and 
two-body meson-cloud diagrams (Figs.7 and 8). Note, that the diagrams in 
Figs.4, 5, and 7 do not contribute to the real part of the spin-independent 
polarizabilities of the nucleon $\alpha_E$ and $\beta_M$. They only contribute 
to the spin-dependent polarizabilities. The intermediate ground state quark 
propagators contribute only in the diagrams of Figs.2 and 8 (both 
to $\alpha_E$ and to $\beta_M$). In the case of the diagrams of Fig.6 the 
intermediate ground quarks state contribute to the magnetic 
polarizabilities only. Excited intermediate quark states contribute to the 
tree-level (Fig.1) and meson-cloud diagrams (Figs.2  and 6). 

For illustration we present the expressions for some typical diagrams 
contributing to the $T^N$ matrix. The contribution of the tree-level 
diagrams (Figs.1a and 1b) is: 
\begin{eqnarray}\label{T1}
T_{1}^N&=&\epsilon_{m}^{\ast\prime} \, I_{1}^{mn} \, \epsilon_{n} 
\,, \hspace*{1cm} I_{1}^{mn} = I_{\rm 1a}^{mn} + I_{\rm 1b}^{mn}\,,
\nonumber \\
I_{\rm 1a}^{mn}&=& e^2 \, \sum\limits_{i=1}^3 \, \sum\limits_{\alpha} \,  
\frac{V_i^{m}(\vec{k}^{\,\prime};\alpha) \, 
V_i^{n \dagger}(\vec{k};\alpha)}
{\Delta E_{\alpha}-\omega-i0^+}\,, ; 
I_{\rm 1b}^{mn} \, = \, e^2 \, \sum\limits_{i=1}^3 \, \sum\limits_{\alpha} \,  
\frac{V_i^{n}(-\vec{k};\alpha) \, 
V_i^{m \dagger}(-\vec{k}^{\,\prime};\alpha)}
{\Delta E_{\alpha}+\omega-i0^+},
\end{eqnarray}
where $(m,n)=1,2,3$, and 
$\Delta E_{\alpha}$ is the energy difference between the intermediate 
quark state (labeled by $\alpha$) and the initial 
(or final) quark ground state. The vertex function of the photon-quark 
coupling in Eq.~(\ref{T1}) is 
\begin{eqnarray}
V_i^{m}(\vec{k};\alpha)=\int d^3x \bar{u}_0(x;i)\gamma^{m}Q_i u_{\alpha}(x;i) 
e^{-i\vec{k} \cdot \vec{x}},
\end{eqnarray}
where $u_0(x;i)$ and $u_{\alpha}(x;i)$ are the wave functions of ground and 
excited quark states; $Q_i$ is the charge number of the $i$-th quark. The 
amplitude $I_1^{mn}$ is proportional to $\sum_{i=1}^3Q_i^2$, resulting in  
1 for proton and $2/3$ for neutron. If only the ground state is 
considered, the three-component of the vertex function is 
\begin{eqnarray}
\vec{V}_i(\vec{k};0)=iN_0^2\rho R (\pi R^2)^{3/2}e^{-\frac{k^2R^2}{4}} \,Q_i\, 
[\vec{\sigma}(i)\times\vec{k}],
\end{eqnarray}
of which the quantities R, $\rho$, and $N_0$ are defined in Sec. II. 
Tree-level diagrams constrained to include only intermediate ground state 
quarks do not contribute to the nucleon polarizabilities. This is due to the 
fact that $\Delta E_{\alpha}=0$ and, therefore, 
the expansion of $T_{1}^{mn}$ in powers of $\omega$ does not contain 
$\omega^2$ terms. Inclusion of the intermediate excited quark states in 
the tree-level diagrams leads to a contribution to the nucleon 
polarizabilities due to $\Delta E_{\alpha} \not = 0$. Moreover, only 
$\beta_M$ is affected by the tree-level diagrams, 
because they have the following spin structure 
\begin{eqnarray*}
\vec{\sigma}(i) \cdot (\vec{k}^{\,\prime} \times 
\vec{\epsilon}^{\,\,\ast\prime} ) \; 
\vec{\sigma}(i) \cdot (\vec{k}\times\vec{\epsilon}) = 
\omega^2 (\cos\theta \, \vec{\epsilon} \cdot \vec{\epsilon}^{\,\,\ast\prime} 
- \vec{\epsilon}^{\,\,\ast\prime} \cdot \hat{k} \, 
\vec{\epsilon}\cdot\hat{k}^\prime) 
\, + \, {\rm spin}-{\rm dependent \ part}\,. 
\end{eqnarray*}

As already mentioned before, only the meson-cloud diagrams of Figs.2, 3, 6, 
and 8 contribute to the spin-independent nucleon polarizabilities $\alpha_E$ 
and $\beta_M$. In other chiral approaches, the one-body diagrams 
of Figs.2 and 3 give rise to the dominant terms to $\alpha_E$ and $\beta_M$. 
In addition, in our approach we take into account two-body quark forces. 
Therefore, $\alpha_E$ and $\beta_M$ receive a contribution from the diagrams 
of Fig.8 which are the two-body analogue of the one-body diagrams of Fig.2. 
The diagrams of Fig.6 evaluated for the nucleon polarizabilities are 
suppressed. For example, the explicit expression of the diagrams in Fig.2 
to the $T^N$ matrix is given by 
\begin{eqnarray}
T_{2}^N&=&\epsilon_{m}^{\ast\prime} \, I_{2}^{mn} \, \epsilon_{n} 
\,, \hspace*{1cm} I_{2}^{mn} = I_{\rm 2a}^{mn} + I_{\rm 2b}^{mn} 
+ I_{\rm 2c}^{mn}\,,
\end{eqnarray}
where 
\begin{eqnarray}
I_{2a}^{mn} + I_{2b}^{mn} &=& - \frac{e^2}{F^2}\sum_{i=1}^3
\, \sum\limits_{\alpha} \,  
\int\frac{d^4q}{(2\pi)^4i}\nonumber \\
&\times&\biggl[ P_{\alpha}(\vec{q}-\vec{k}';i)\Delta_{\pi}(q-k')
\Delta_{\pi}(q)\Delta_{\pi}(q-k)
P_{\alpha}^\dagger(\vec{q}-\vec{k};i)\frac{(2q-k')^m(2q-k)^n}
{q_0+\Delta E_\alpha-\omega-i0^+} \nonumber\\
&+&P_{\alpha}(\vec{q}+\vec{k};i)\Delta_{\pi}(q+k')
\Delta_{\pi}(q)\Delta_{\pi}(q+k)
P_{\alpha}^\dagger(\vec{q}+\vec{k}';i)\frac{(2q+k')^m(2q+k)^n}
{q_0+\Delta E_\alpha+\omega-i0^+} \biggr]\,,\nonumber\\
& &\\
I_{2c}^{mn}&=& - \delta^{mn} \frac{e^2}{F^2}\sum_{i=1}^3
\, \sum\limits_{\alpha} \,  
\int\frac{d^4q}{(2\pi)^4i}\, 
\frac{1}{q_0+\frac{\Delta k_0}{2}+\Delta E_\alpha-i0^+}\nonumber \\
&\times&P_{\alpha}\biggl(\vec{q}-\frac{\Delta \vec{k}}{2};i\biggr)
\Delta_{\pi}\biggl(q-\frac{\Delta k}{2}\biggr)
\Delta_{\pi}\biggl(q+\frac{\Delta k}{2}\biggr)
P_{\alpha}^\dagger\biggl(\vec{q}+\frac{\Delta \vec{k}}{2};i\biggr)\nonumber
\end{eqnarray}
with $q$ being the pion momentum and $\Delta k=k'-k$. 
Here, $P_{\alpha}(\vec{k};i)$ is the quark-pion vertex function 
\begin{eqnarray}
P_{\alpha}(\vec{k};i)&=&\int d^3x \bar{u}_0(x;i)\gamma_5 S(r) 
u_{\alpha}(x;i)e^{i\vec{k}\cdot\vec{x}} 
\end{eqnarray}
where $S(r)$ is the scalar part of the confinement potential. 
In particular, when $\alpha=0$,  we have 
\begin{eqnarray}
P_0(\vec{k};i)&=&-\frac{3}{10} \, g_A \, F_{\pi NN}(\vec{k}^2) \, 
\vec{\sigma}(i)\cdot\vec{k}\,,\\
F_{\pi NN}(\vec{k}^{\,2})& = & 
\exp\biggl(-\frac{\vec{k}^{\,2} R^2}{4}\biggr)  \, 
\biggl[1+\frac{\vec{k}^{\,2} R^2}{8}\biggl(1-\frac{5}{3 g_A}\biggr)\biggr],  
\nonumber 
\end{eqnarray} 
where $F_{\pi NN}(\vec{k}^{\,2})$ is the $\pi NN$ 
form factor normalized to unity at zero recoil, $\vec{k}^{\,2} = 0$.

Our results for the proton and neutron spin-independent polarizabilities 
$\alpha_E$ and $\beta_M$ are contained in Figs.9-18. In Figs.9-12 we show 
the sensitivity of the proton (Figs.9 and 11) and neutron polarizabilities 
(Figs.10 and 12) to a variation in the model parameter $<r^2_E>^p_{\rm LO}$. 
In Figs.13-18 we demonstrate the contribution of the individual excited 
states in the tree-level (Figs.13 and 14) and in the meson-cloud 
(Figs.15-18) diagrams to $\beta_M^p$ and $\beta_M^n$. The contribution 
of the tree-level diagrams  to $\beta_M$ grows linearly when the size 
parameter $<r^2_E>^p_{\rm LO}$ increases. Results for the meson-cloud 
diagrams are not changed so much as function of $<r^2_E>^p_{\rm LO}$ 
(when we restrict to the inclusion of the ground state quark propagators 
only) and grows (decreases) in $\alpha_E$ ($\beta_M$) when 
$<r^2_E>^p_{\rm LO}$ increases. 

Now we discuss the numerical results for the nucleon polarizabilities 
in the case of the central value $<r^2_E>^p_{\rm LO}=0.6$ fm$^2$ 
obtained  in previous phenomenology~\cite{Lyubovitskij:2001nm,Lyubovitskij:2000sf,Cheedket:2002ik}. 
In Table I we compare our results to the data and to other theories 
(ChPT~\cite{Bernard:1991rq}, ChPT + dispersion relation (DR) 
constraints~\cite{Lvov:1993ex}). In Table II we present the individual 
contributions of the different diagrams to the nucleon polarizabilities, 
while also listing the separate results for the ground and excited quark 
states in the loops. It can be seen that the tree-level diagrams and 
the meson cloud diagrams of Fig. 6 do not contribute to the electric 
polarizabilities. The contribution of the tree-level diagrams to the 
magnetic polarizabilities  is solely due to the inclusion of the excited quark 
states. As already indicated, the excited state contributions in the 
tree-level diagrams are for the case of the neutron 2/3 of the result 
of the proton. The total predicted value for $\beta_M^p$ is larger than that 
of $\beta^n$.  Finally, we found that only very small differences appear 
in the contribution of the six diagrams of Fig.6 to $\beta_M$, which changes  
slightly from 0.085 for the proton to 0.089 for the neutron.
In our model calculations there is no remarkable difference between the proton 
and neutron cases for the pion-cloud diagrams. 
Thus, our results show that $\alpha_E^p=\alpha_E^n$. 
Moreover, the large difference between proton and neutron arises for 
$\beta_M$, where the tree-level diagrams of intermediate excited quark states 
differ strongly for both cases.

\section{CONCLUSIONS}

In this work we apply the perturbative chiral quark model to the description 
of the nucleon spin-independent polarizabilities $\alpha_E$ and $\beta_M$. 
It has been previously verified that the model is successful in the 
explanation of many aspects of nucleon properties [14-18], such as the 
magnetic moments, the axial vector form factor,  the $N \to \Delta$ transition 
amplitude, the meson-nucleon sigma-term and $\pi N$ nucleon scattering. Here, 
we find that the model can reasonably reproduce the electric polarizabilities 
of proton and neutron without the use of any additional free parameters. 
The magnetic polarizabilities, in particular for the proton, are overestimated 
when compared to the present data. This effect is known from experience of 
other theoretical calculations. 

In the perturbative chiral quark model it is found that without excited 
quark states in the loops already reasonable value for $\alpha_E$ and 
a negative one for $\beta_M$ are obtained. Then $\beta_M$ behaves like 
a diamagnetic polarizability. 
This finding also appears in the dispersion relation analysis, 
in other chiral quark model calculations and in leading-order relativistic 
chiral perturbative theory (see discussion in Ref.~\cite{Schumacher:2005an}). 
When we include the excite states in the quark propagators we get 
a paramagnetic polarizability $\beta_M$ arising from the tree-level diagrams.  
Our value for $\beta_M^p$ is overestimated in comparison with the data. 
This trend is also seen in heavy baryon chiral perturbative theory with the 
missing $\Delta(1232)$ contribution in terms of a ``small scale expansion''. 
From a detailed analysis we find that the intermediate excited quark states 
are important to generate a positive value for $\beta_M$. In fact, the 
sizeable contribution of excited states has already been elaborated in our 
calculation of the $\Delta \to N$ transition amplitude 
and in the case of the nucleon axial 
charge~\cite{Cheedket:2002ik}. Comparing each contribution of the five 
intermediate excited states, 
we find that the effects of $1p_{3/2}$ and $1p_{1/2}$ 
are more pronounced than the effects of the other three resonances. 
For the neutron case, the main contribution, to the paramagnetic 
polarizability $\beta_M$ from the excited states of the Born terms, is 
proportional to $\sum_{j=1}^3Q_j^2$. Thus, the paramagnetic polarizability 
of the neutron is smaller than that of the proton target. For the observable 
$\alpha_E$, the one-pion loop contributions,  both from the ground state and 
from the excited states, are not sensitive to isospin. Consequently, we have 
$\alpha^p_E\sim\alpha_E^n$ in the present calculation. 
  
\section*{Acknowledgments}

This work was supported by the  DFG under contracts FA67/25-3 and GRK683. 
This research is also part of the EU Integrated Infrastructure Initiative 
Hadron physics project under contract number RII3-CT-2004-506078 and  
the President grant of Russia "Scientific Schools"  No. 1743.2003. 
The support from the National Natural Science Foundations of China is also 
appreciated. K.P. thanks the Development and Promotion of Science and 
Technology Talent Project (DPST), Thailand for financial support.  
Y.Dong and P.Shen thank the Institut f\"ur Theoretische Physik, 
Universit\"at T\"ubingen, for their hospitality. 

\newpage

\section*{Appendix A. 
Solutions of the Dirac equation for the effective potential}
\par\noindent\par\noindent

In this section we indicate the solutions to the Dirac equation
with the effective potential $V_{\rm eff}(r) = S(r) + \gamma^0
V(r)$. The scalar $S(r)$ and time-like vector $V(r)$ parts are
given by
\begin{eqnarray}
S(r)&=& M_1 + c_1 r^2,
\nonumber \\
V(r) &=& M_2 + c_2 r^2,
\end{eqnarray}
with the particular choice
\begin{eqnarray}
M_1 = \frac{1 \, - \, 3\rho^2}{2 \, \rho R} , \hspace*{1cm}
M_2 = {\cal E}_0 - \frac{1 \, + \, 3\rho^2}{2 \, \rho R} , \hspace*{1cm}
c_1 \equiv c_2 =  \frac{\rho}{2R^3} .
\end{eqnarray}
The quark wave function $u_{\alpha}(\vec r)$ in state $\alpha$
with eigen-energy ${\cal E}_{\alpha}$ satisfies the Dirac equation
\begin{equation}
[-i \vec \alpha \vec \nabla +\beta S(r) + V(r) - {\cal E}_{\alpha}]
u_{\alpha} (\vec r) = 0.
\end{equation}
Solutions of the Dirac spinor $u_{\alpha}(\vec r)$ can
be written in the form~\cite{Tegen:1982tg}
\begin{equation}
u_{\alpha}(\vec r)= N_{\alpha}
\left( \begin{array}{c} g_{\alpha}(r) \\ i \vec \sigma
\cdot \hat r f_{\alpha}(r) \end{array} \right) {\cal Y}_{\alpha}(\hat r)
\chi_f \chi_c .
\end{equation}
For the particular choice of the potential the radial functions 
$g$ and $f$ satisfy the form
\begin{equation}
g_{\alpha}(r) = \bigg( \frac{r}{R_{\alpha}} \bigg)^l
L^{l+1/2}_{n-1}\bigg( \frac{r^2}{R^2_{\alpha}} \bigg) e^{-\frac{r^2}{2
R^2_{\alpha}}},
\end{equation}
where for $j=l+\frac{1}{2}$
\begin{equation}
f_{\alpha}(r)=\rho_{\alpha} \bigg(\frac{r}{R_{\alpha}}\bigg)^{l+1}
\bigg[L^{l+3/2}_{n-1}(\frac{r^2}{R^2_{\alpha}}) +
L^{l+3/2}_{n-2}(\frac{r^2}{R^2_{\alpha}})\bigg] 
e^{-\frac{r^2}{2 R^2_{\alpha}}},
\end{equation}
and for $j=l-\frac{1}{2}$
\begin{equation}
f_{\alpha}(r)=-\rho_{\alpha} \bigg(\frac{r}{R_{\alpha}}\bigg)^{l-1}
\bigg[(n+l-\frac{1}{2})L^{l-1/2}_{n-1}(\frac{r^2}{R^2_{\alpha}})
+ nL^{l-1/2}_{n}(\frac{r^2}{R^2_{\alpha}})\bigg] e^{-\frac{r^2}
{2 R^2_{\alpha}}}.
\end{equation}
The label $\alpha=(nljm)$ characterizes the state with
principle quantum number $n=1,2,3,...$, orbital angular momentum $l$,
total angular momentum $j=l\pm \frac{1}{2}$ and projection $m$.
Due to the quadratic nature of the potential the radial
wave functions contain the associated
Laguerre polynomials $L^{k}_{n}(x)$ with
\begin{equation}
L^{k}_{n}(x)=\sum^{n}_{m=0} (-1)^m \frac{(n+k)!}{(n-m)!(k+m)!m!} x^m.
\end{equation}
The angular dependence (${\cal Y}_{\alpha}(\hat r) \equiv
{\cal Y}_{lmj}(\hat r)$) is defined by
\begin{equation}
{\cal Y}_{lmj}(\hat r)=\sum_{m_l,m_s} (l m_l \frac{1}{2} m_s | j m)
Y_{l m_l}(\hat r) \chi_{\frac{1}{2} m_s}
\end{equation}
where $Y_{l m_l}(\hat r)$ is the usual spherical harmonic.
Flavor and color parts of the Dirac spinor are represented
by $\chi_f$ and $\chi_c$, respectively.

The normalization constant is obtained from the condition
\begin{equation}
\int\limits^\infty_0 d^3 \vec r u^{\dagger}_{\alpha}(\vec r)
u_{\alpha}(\vec r) = 1
\end{equation}
which results in
\begin{equation}
N_{\alpha}=\bigg[ 2^{-2(n+l+1/2)} \pi^{1/2} R^3_{\alpha}
\frac{(2n+2l)!}{(n+l)!(n-1)!}\{1 + \rho^2_{\alpha}
(2n + l -\frac{1}{2})\} \bigg]^{-1/2},
\end{equation}
The two coefficients $R_{\alpha}$ and $\rho_{\alpha}$ are of
the form
\begin{eqnarray}
R_{\alpha} &=& R(1 + \Delta {\cal E}_{\alpha} \rho R)^{-1/4},\\
\rho_{\alpha} &=& \rho \bigg( \frac{R_{\alpha}}{R} \bigg)^3,
\end{eqnarray}
and are related to the Gaussian parameters $\rho$ and  $R$ 
of Eq.~(\ref{eigenstate}). 
The quantity $\Delta {\cal E}_{\alpha} ={\cal E}_{\alpha}- {\cal E}_0$
is the difference between the energy
of state $\alpha$ and the ground state.
$\Delta {\cal E}_{\alpha}$ depends on the quantum numbers $n$ and $l$
and is related to the parameters $\rho$ and $R$ by
\begin{equation}
(\Delta {\cal E}_{\alpha} + \frac{3 \rho}{R})^2
(\Delta {\cal E}_{\alpha} + \frac{1}{\rho R}) =
\frac{\rho}{R^3} (4 n +2 l -1)^2.
\end{equation}

\newpage
\begin{figure}[t]

\noindent Fig.1: Tree-level diagrams 
contributing to Compton scattering on a quark.  

\vspace*{.1cm}

\noindent Figs.2-6: One-body meson-cloud diagrams in Compton scattering.

\vspace*{.1cm}

\noindent Figs.7-8: Two-body meson-cloud diagrams in Compton scattering.

\vspace*{.1cm}

\noindent Fig.9: 
Proton electric polarizability $\alpha_E^p$: The dotted and the dashed 
curves are the results of the pion loops without and with inclusion of the 
excited states. The total value is shown by the solid line. The shaded 
region shows the experimental results including error 
bars~\cite{Schumacher:2005an}. 

\vspace*{.1cm}

\noindent Fig.10: 
Neutron electric polarizability $\alpha_E^n$: Otherwise as in Fig.9.

\vspace*{.1cm}

\noindent Fig.11: Proton magnetic polarizability $\beta_M^p$: 
The dotted and dashed curves describe the results of the pion loops 
without and with inclusion of the excited states, respectively. 
The dotted-dashed line is the contribution from the tree-level diagrams 
(excited intermediate quark states). The total value is shown by the solid 
line. The shaded region shows the experimental results including error 
bars~\cite{Schumacher:2005an}. 

\vspace*{.1cm}

\noindent Fig.12: Neutron magnetic polarizability $\beta_M^n$: 
Otherwise as in Fig.11. 

\vspace*{.1cm}
 
\noindent Fig.13: Contribution of the tree-level diagrams (Fig.1) 
[excited intermediate quark states] to $\beta_M^p$. 
The dotted, dashed, dotted-dashed, two-dotted-dashed and two-dashed-dotted 
lines stand for the results of the individual excited states $1p_{1/2}$, 
$1p_{3/2}$, $1d_{3/2}$, $1d_{5/2}$ and $2s_{1/2}$. 

\vspace*{.1cm}
 
\noindent Fig.14: Contribution of the tree-level diagrams (Fig.1) 
[excited intermediate quark states] to $\beta_M^n$. Otherwise as in Fig.13.

\vspace*{.1cm}

\noindent Fig.15: Contribution of the meson-cloud diagram (Fig.2) 
[excited intermediate quark states] to $\alpha_E^p$. 
The dotted, dashed, dotted-dashed, two-dotted-dashed and two-dashed-dotted 
lines stand for the results of the individual excited states $1p_{1/2}$, 
$1p_{3/2}$, $1d_{3/2}$, $1d_{5/2}$ and $2s_{1/2}$. 

\vspace*{.1cm}

\noindent Fig.16: Contribution of the meson-cloud diagram (Fig.2) 
[excited intermediate quark states] to $\alpha_E^n$. Otherwise as in Fig.15.

\vspace*{.1cm}

\noindent Fig.17: Contribution of the meson-cloud diagram (Fig.2) 
[excited intermediate quark states] to  $\beta_M^p$. 
The dotted, dashed, dotted-dashed, two-dotted-dashed and two-dashed-dotted 
lines stand for the results of the individual excited states $1p_{1/2}$, 
$1p_{3/2}$, $1d_{3/2}$, $1d_{5/2}$ and $2s_{1/2}$. 

\vspace*{.1cm}

\noindent Fig.18: Contribution of the meson-cloud diagram (Fig.2) 
[excited intermediate quark states] to  $\beta_M^n$. 
Otherwise as in Fig.17.

\end{figure}

\newpage

\begin{figure}

\vspace*{-2cm}

\hspace*{4cm}
\centering{\
\epsfig{figure=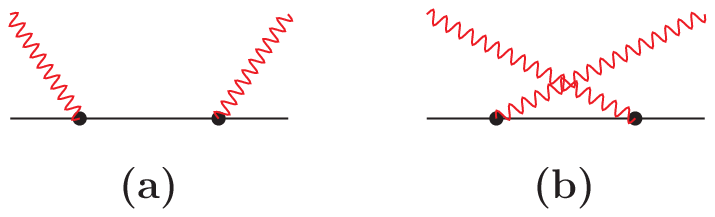,height=12cm}}

\vspace*{-7cm}

\centerline{\bf Fig.1}
\end{figure}

\vspace*{1cm} 

\begin{figure}
\hspace*{.25cm}
\centering{\
\epsfig{figure=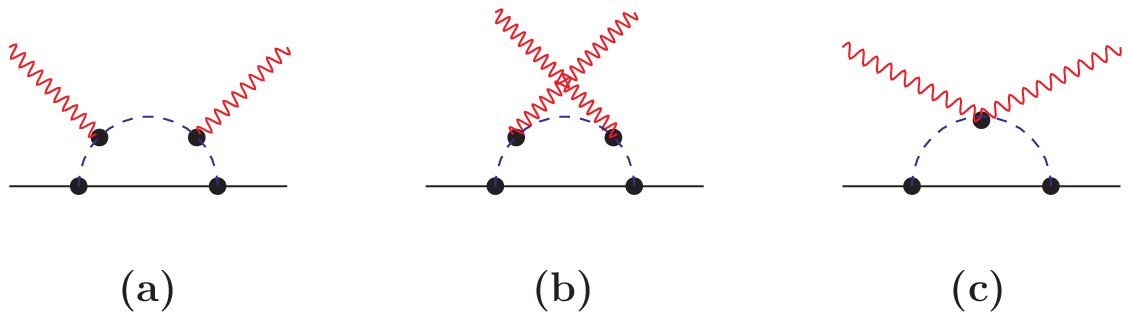,height=12cm}}

\vspace*{-8.5cm}

\centerline{\bf Fig.2}
\end{figure}

\vspace*{1cm} 

\begin{figure}
\hspace*{.25cm}
\centering{\
\epsfig{figure=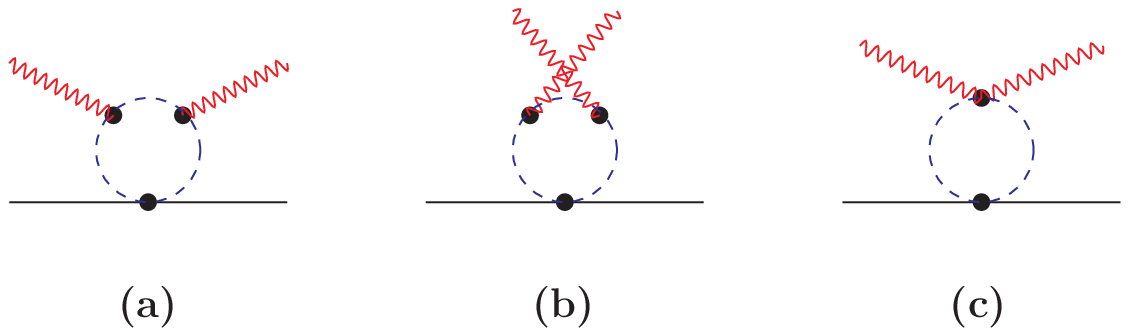,height=12cm}}

\vspace*{-8.5cm}

\centerline{\bf Fig.3}
\end{figure}

\vspace*{1cm} 

\begin{figure}
\hspace*{-2.25cm}
\centering{\
\epsfig{figure=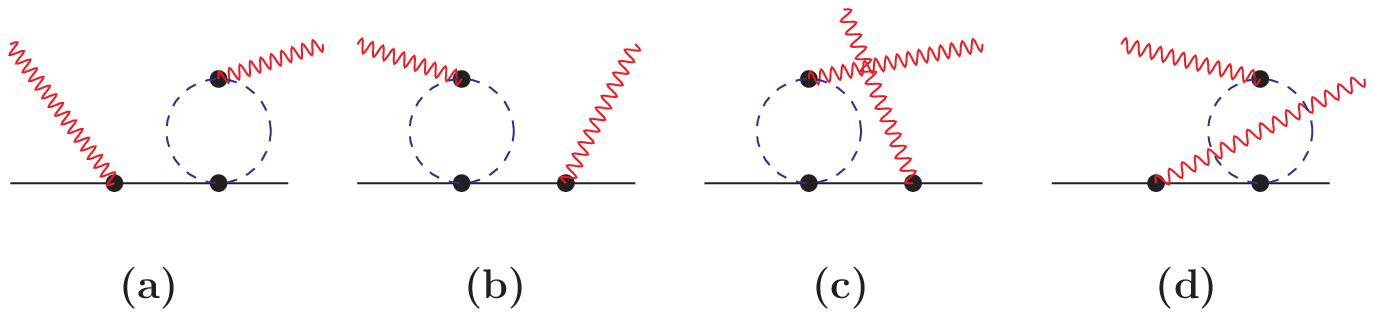,height=12cm}}

\vspace*{-8.5cm}

\centerline{\bf Fig.4}
\end{figure}

\newpage 

\begin{figure} 
\hspace*{.25cm}
\centering{\
\epsfig{figure=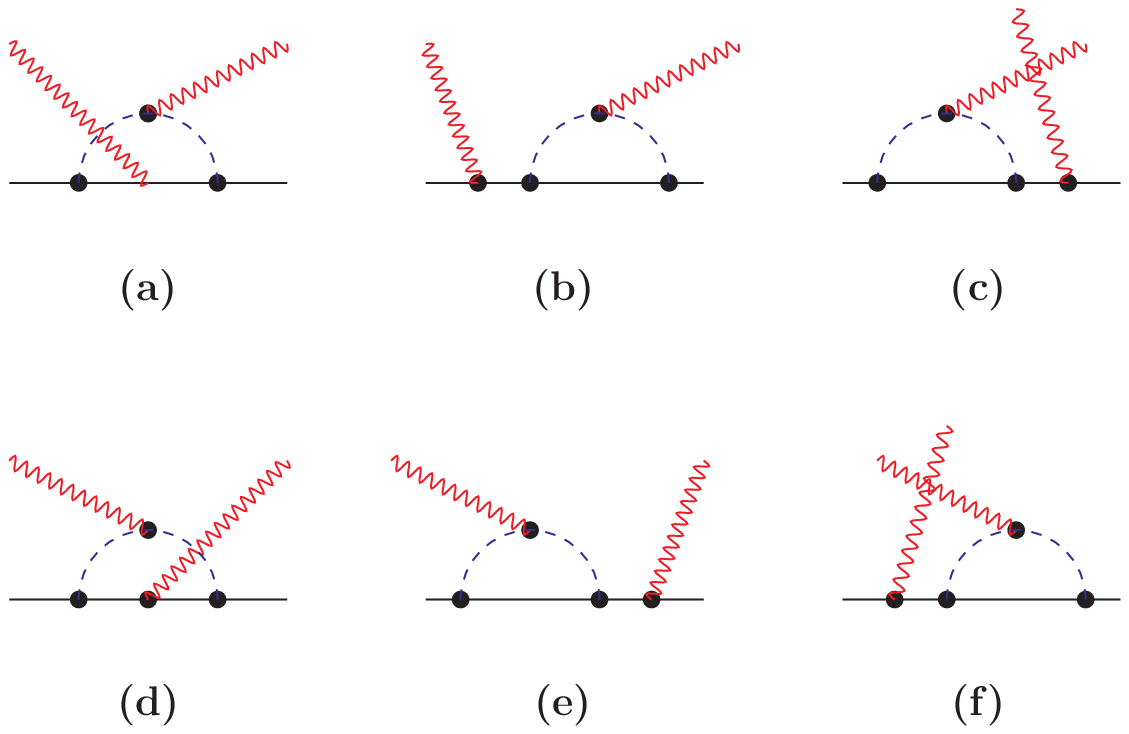,height=12cm}}

\vspace*{-4cm}

\centerline{\bf Fig.5}
\end{figure}

\vspace*{1cm} 

\begin{figure}
\hspace*{.25cm}
\centering{\
\epsfig{figure=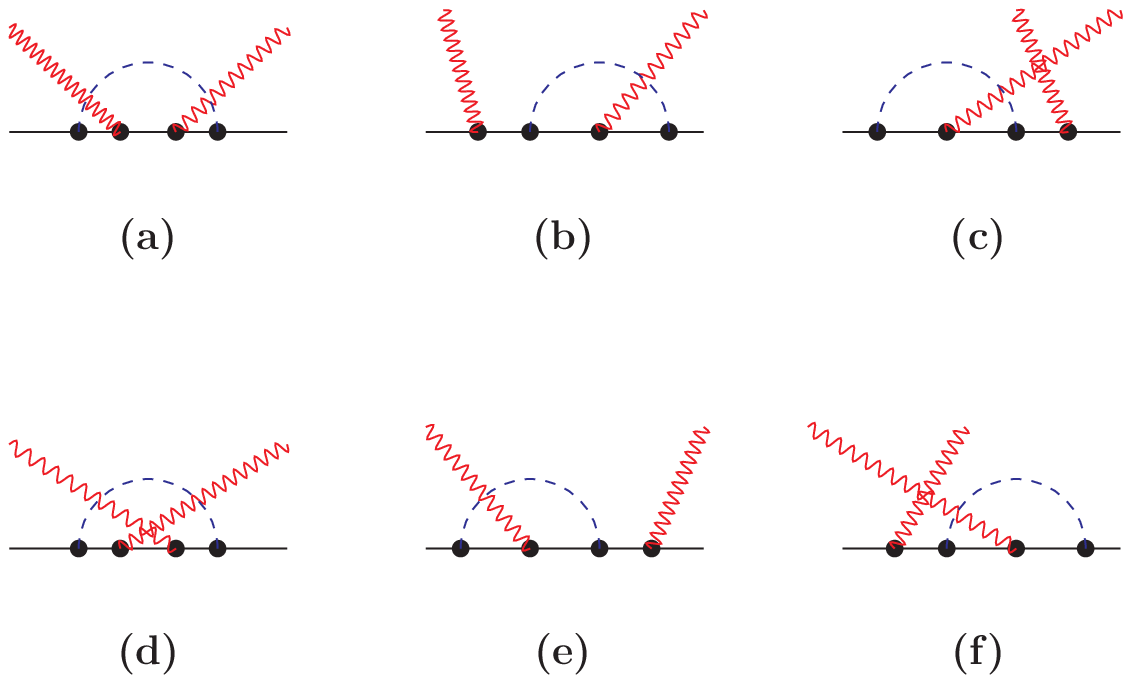,height=12cm}}

\vspace*{-4cm}

\centerline{\bf Fig.6}
\end{figure}

\newpage 

\begin{figure} 
\hspace*{-1.75cm}
\centering{\
\epsfig{figure=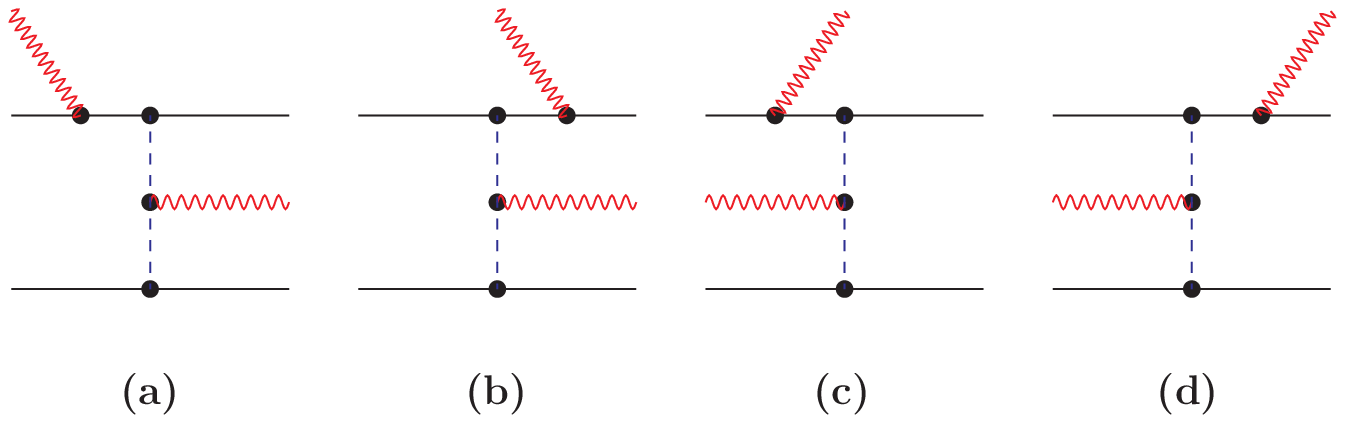,height=12cm}}

\vspace*{-6.5cm}

\centerline{\bf Fig.7}
\end{figure}

\begin{figure}
\hspace*{.25cm}
\centering{\
\epsfig{figure=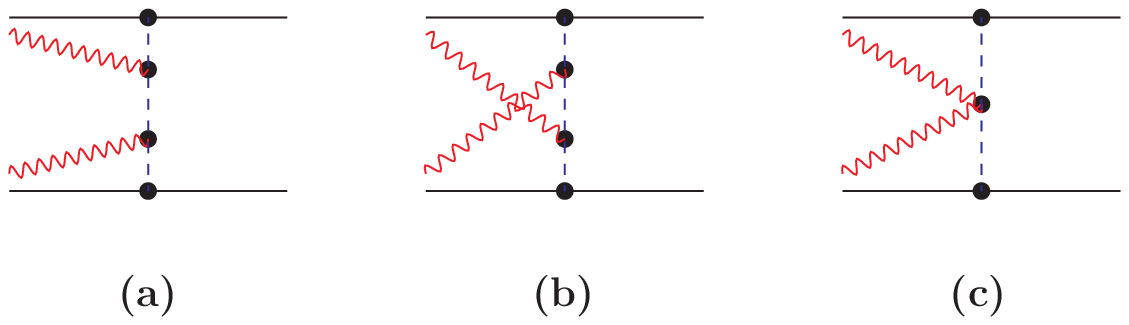,height=12cm}}

\vspace*{-6cm}

\centerline{\bf Fig.8}
\end{figure}

\newpage 

\vspace*{1cm}

\begin{figure}
\centering{\
\epsfig{figure=fig9.eps,scale=0.40}} 

\vspace*{.45cm}

\centerline{\bf Fig.9}
\end{figure}


\vspace*{3cm}

\begin{figure}
\centering{\
\epsfig{figure=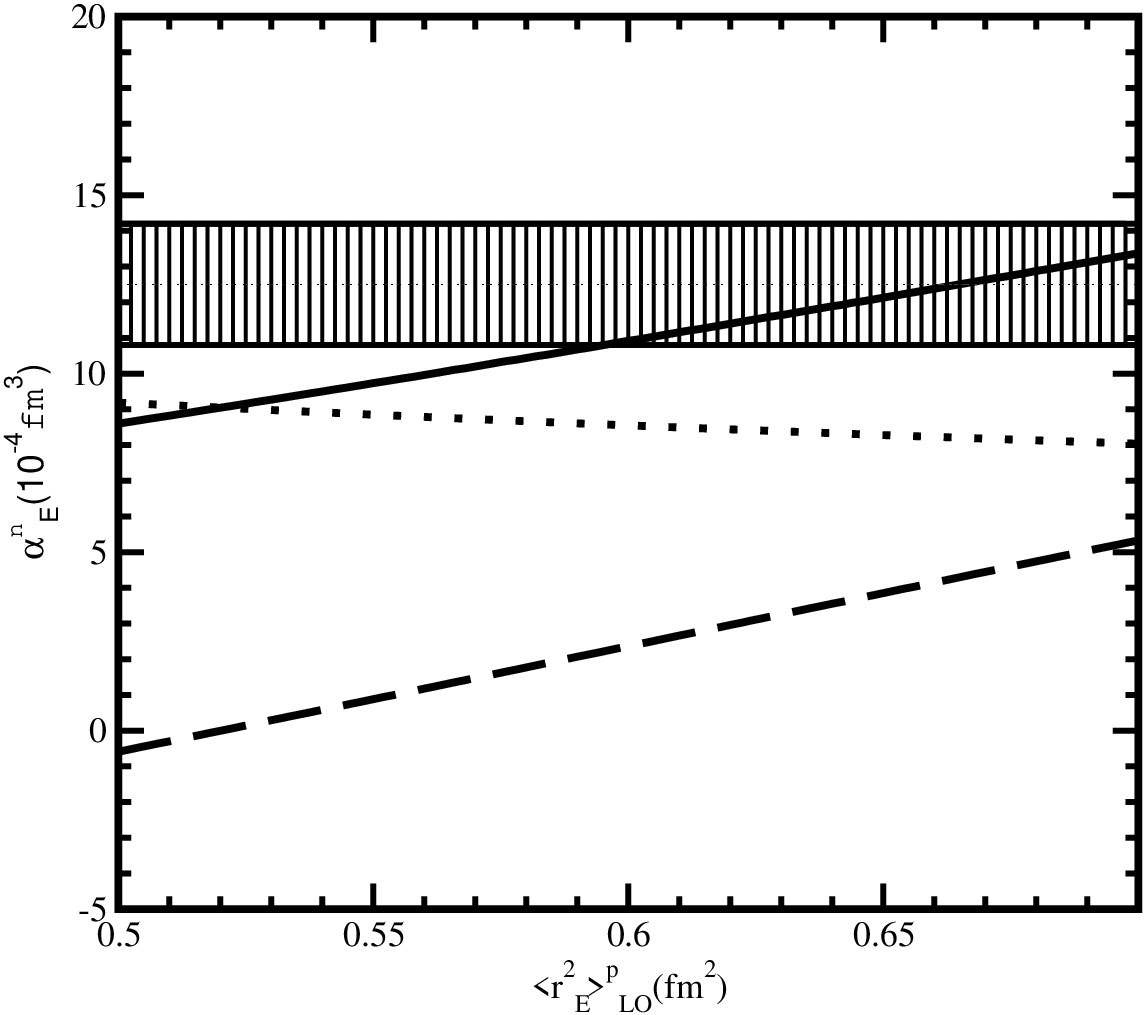,scale=0.40}} 

\vspace*{.45cm}

\centerline{\bf Fig.10}
\end{figure}

\newpage 

\vspace*{1cm}

\begin{figure}
\centering{\
\epsfig{figure=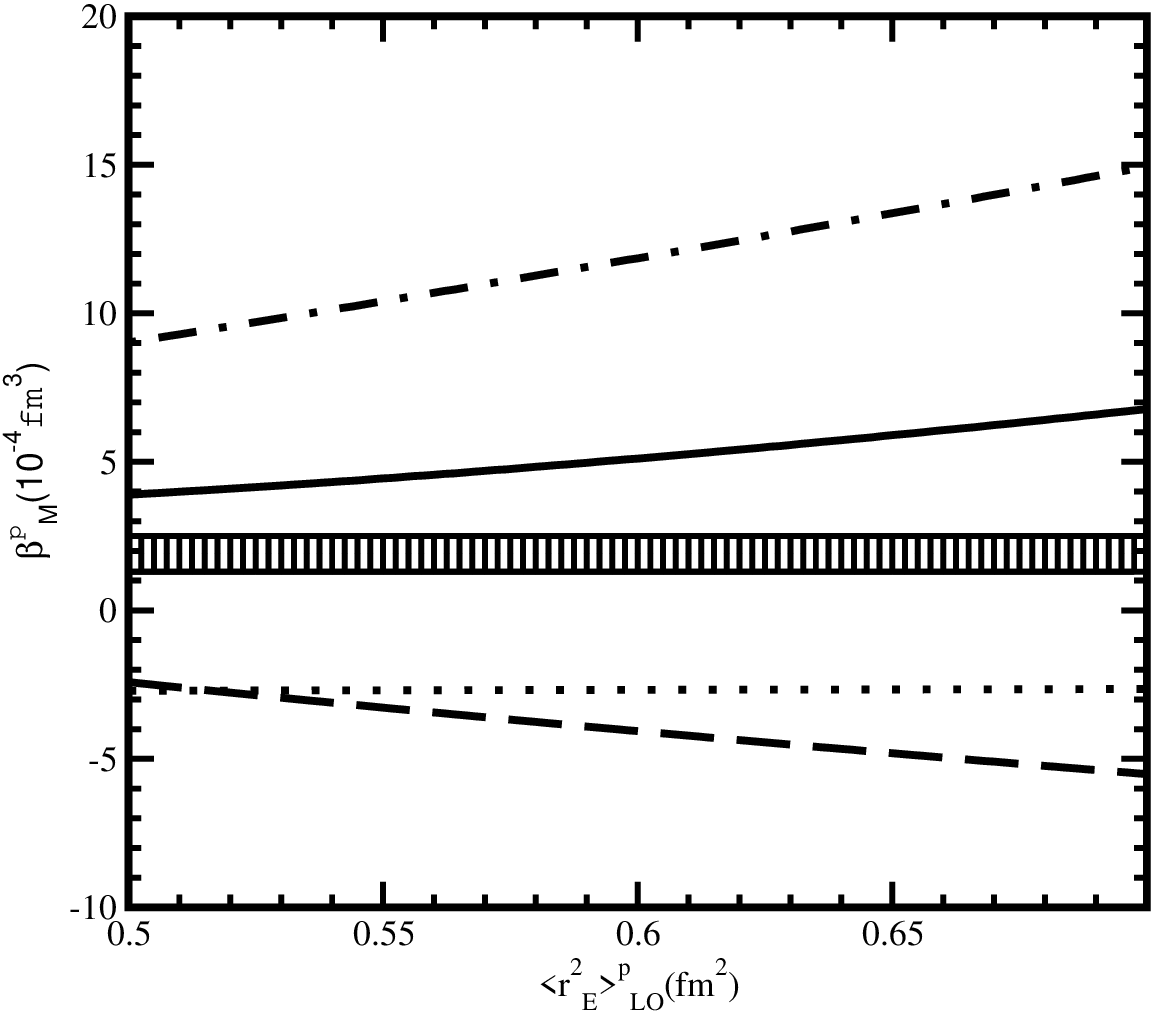,scale=0.40}} 

\vspace*{.45cm}

\centerline{\bf Fig.11}
\end{figure}


\vspace*{3cm}

\begin{figure}
\centering{\
\epsfig{figure=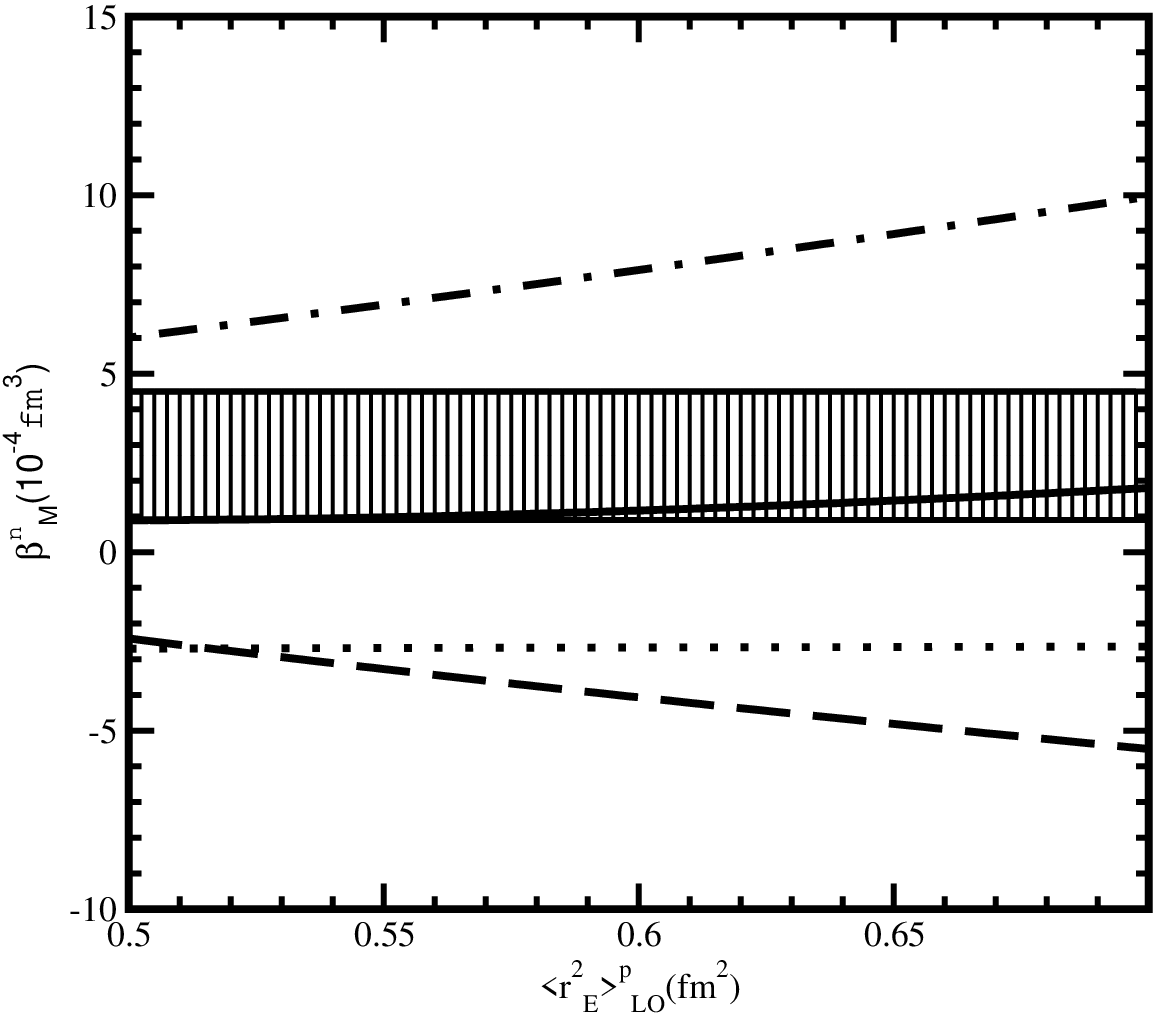,scale=0.40}} 

\vspace*{.45cm}

\centerline{\bf Fig.12}
\end{figure}

\newpage 

\vspace*{1cm}

\begin{figure}
\centering{\
\epsfig{figure=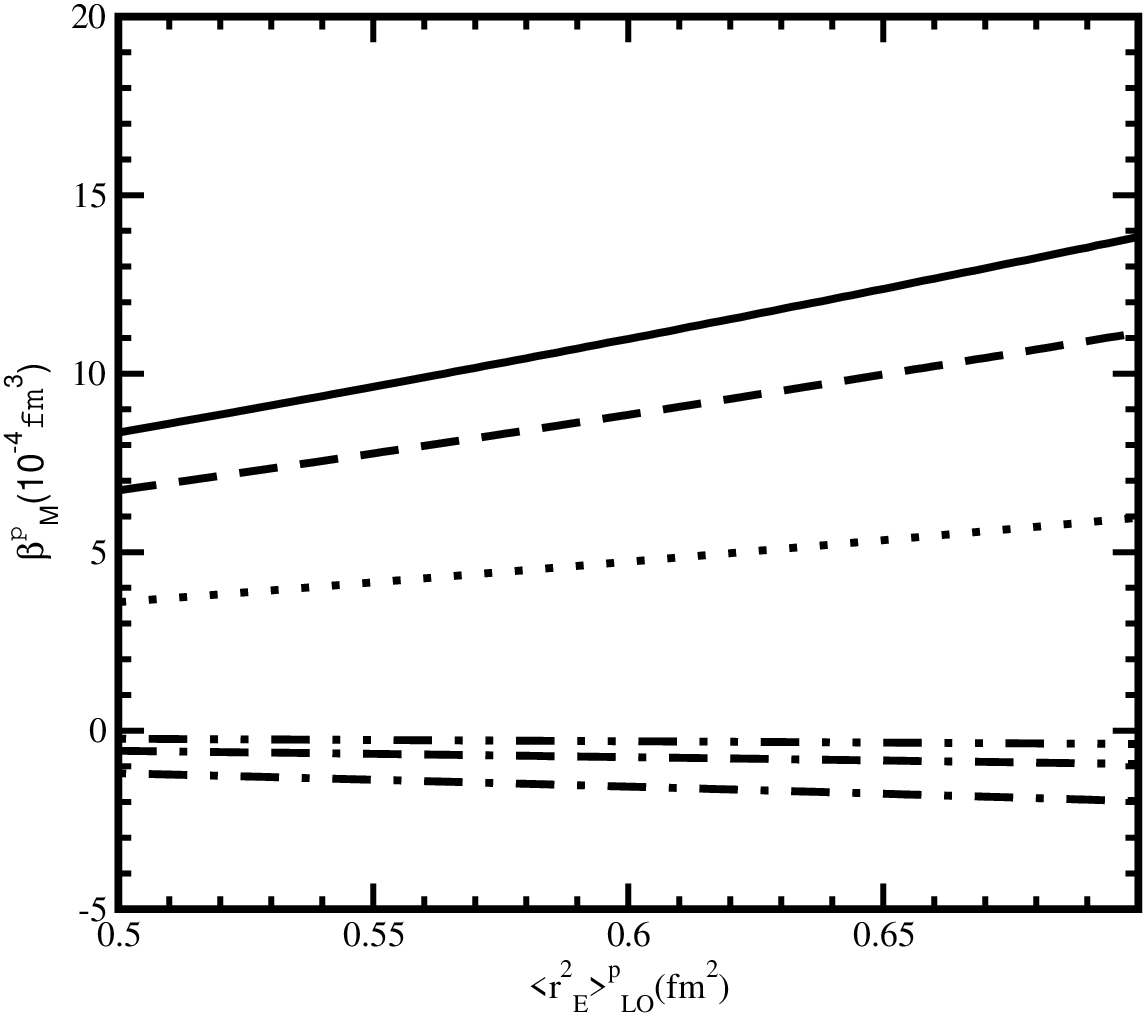,scale=0.40}} 

\vspace*{.45cm}

\centerline{\bf Fig.13}
\end{figure}


\vspace*{3cm}

\begin{figure}
\centering{\
\epsfig{figure=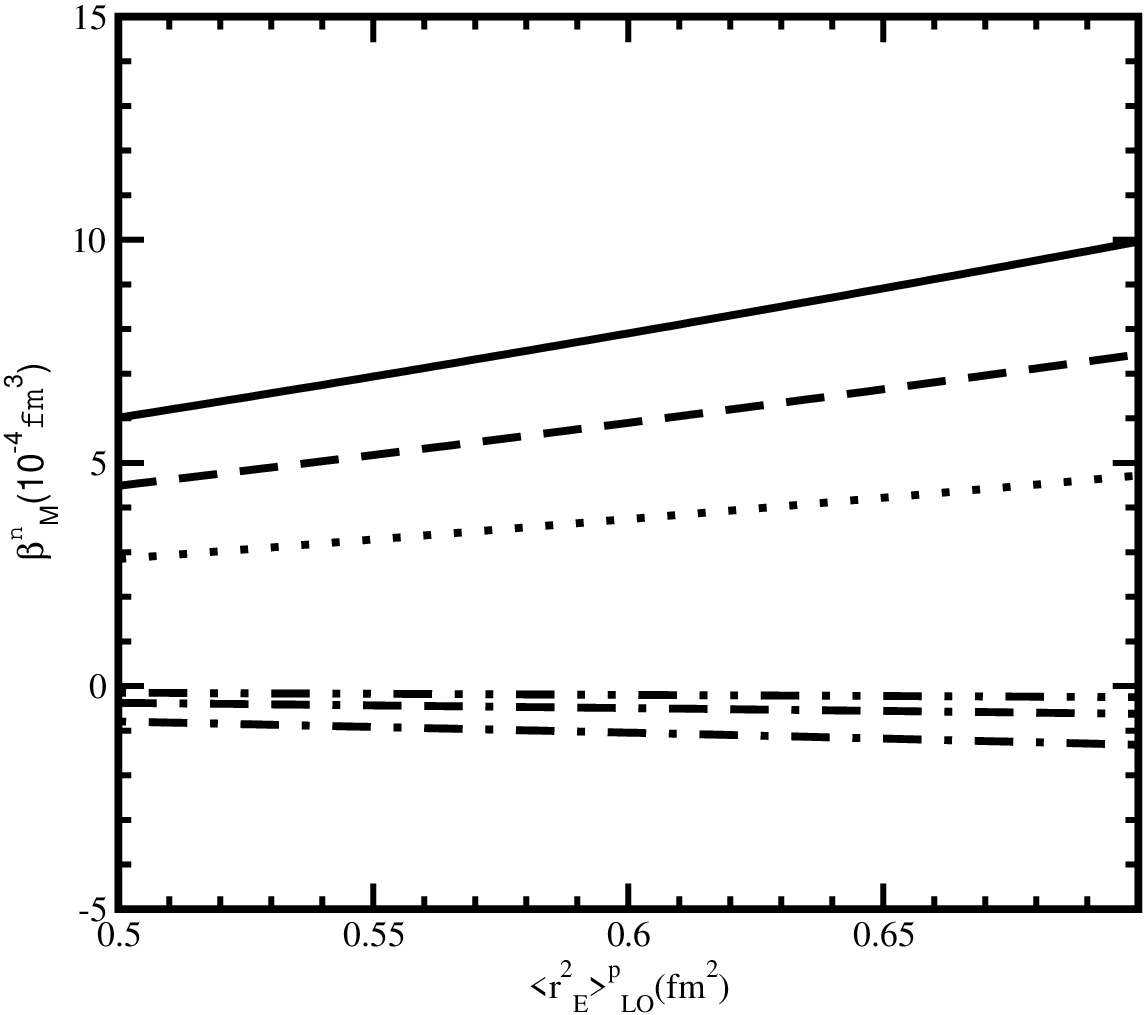,scale=0.40}} 

\vspace*{.45cm}

\centerline{\bf Fig.14}
\end{figure}

\newpage 

\vspace*{1cm}

\begin{figure}
\centering{\
\epsfig{figure=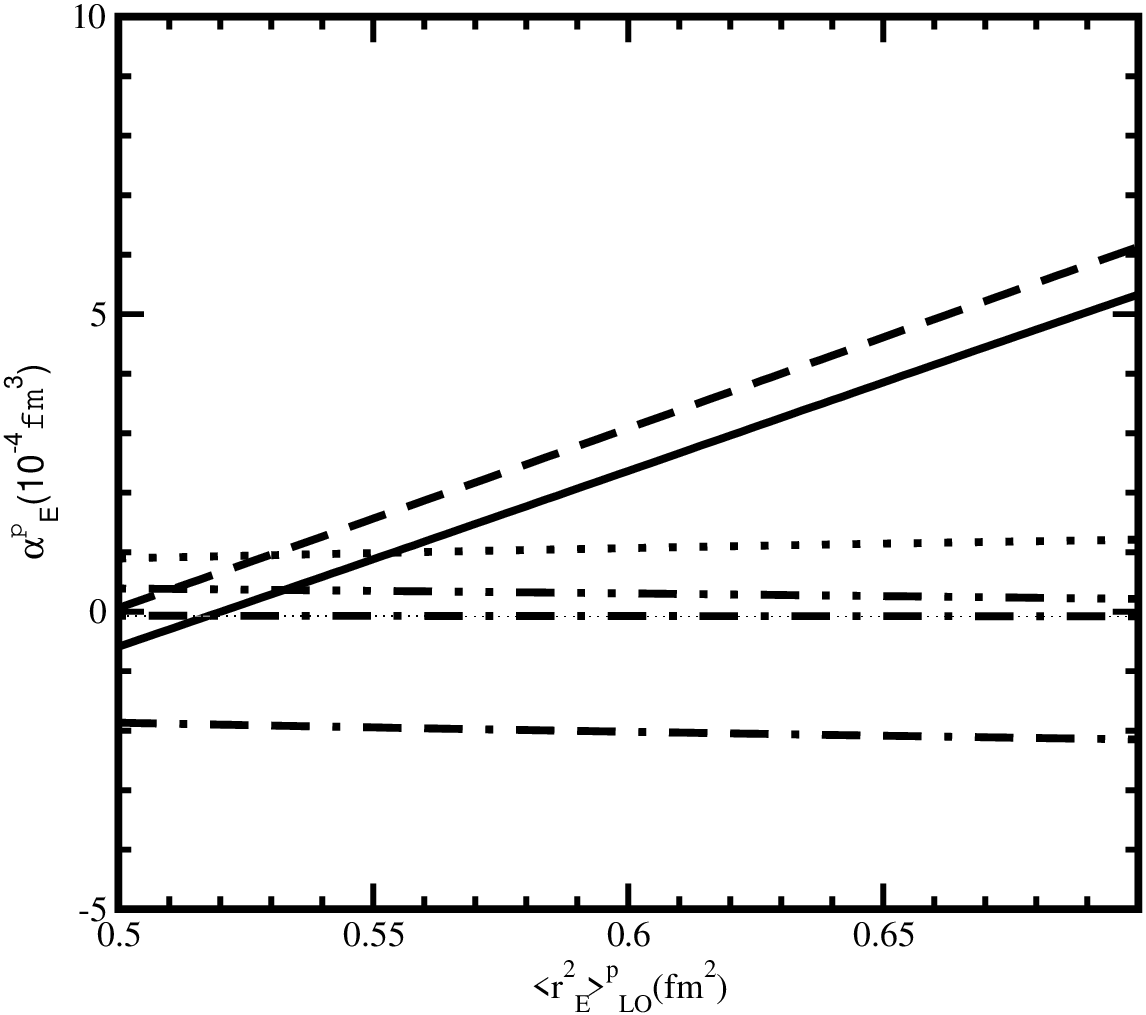,scale=0.40}} 

\vspace*{.45cm}

\centerline{\bf Fig.15}
\end{figure}


\vspace*{3cm}

\begin{figure}
\centering{\
\epsfig{figure=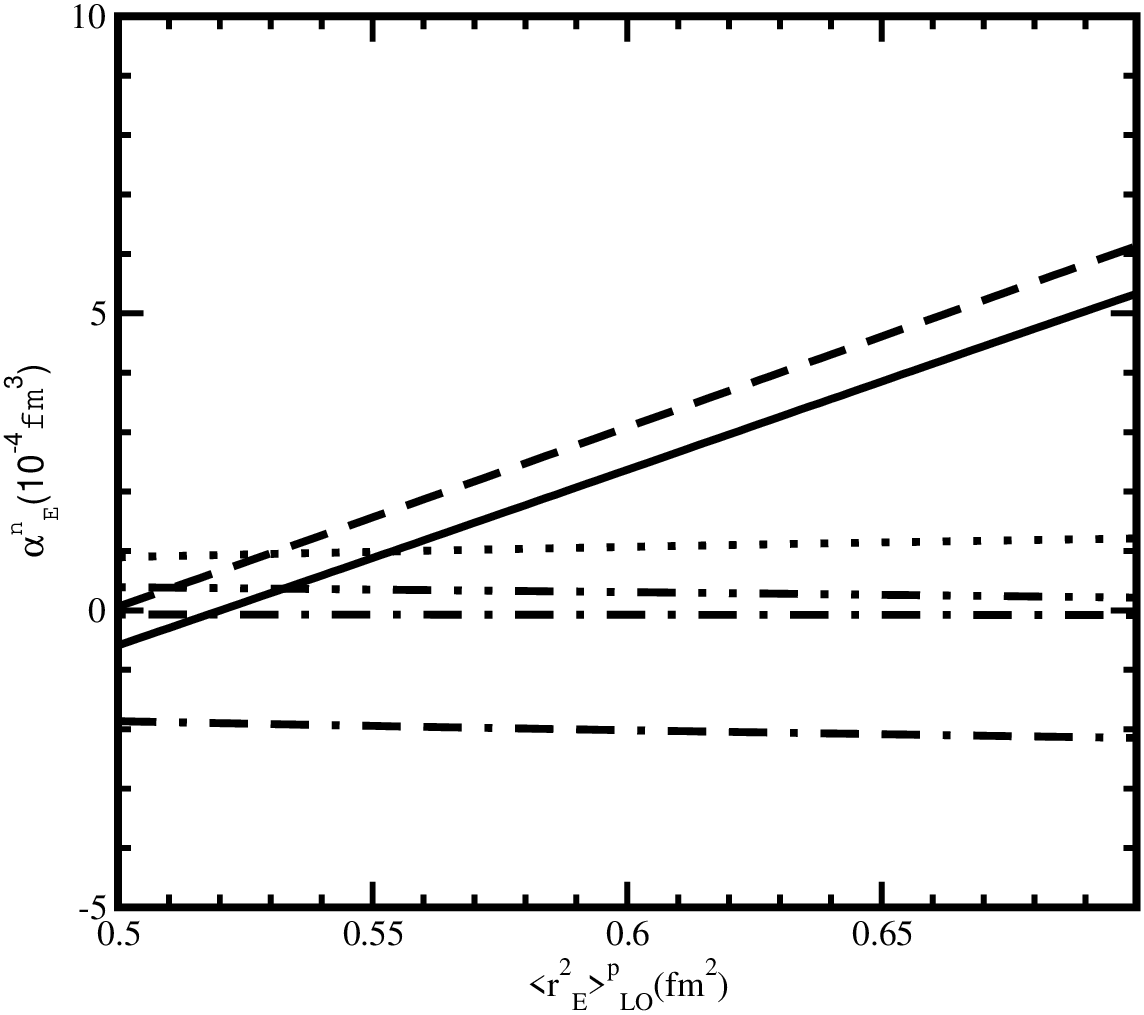,scale=0.40}} 

\vspace*{.45cm}

\centerline{\bf Fig.16}
\end{figure}

\newpage 

\vspace*{1cm}

\begin{figure}
\centering{\
\epsfig{figure=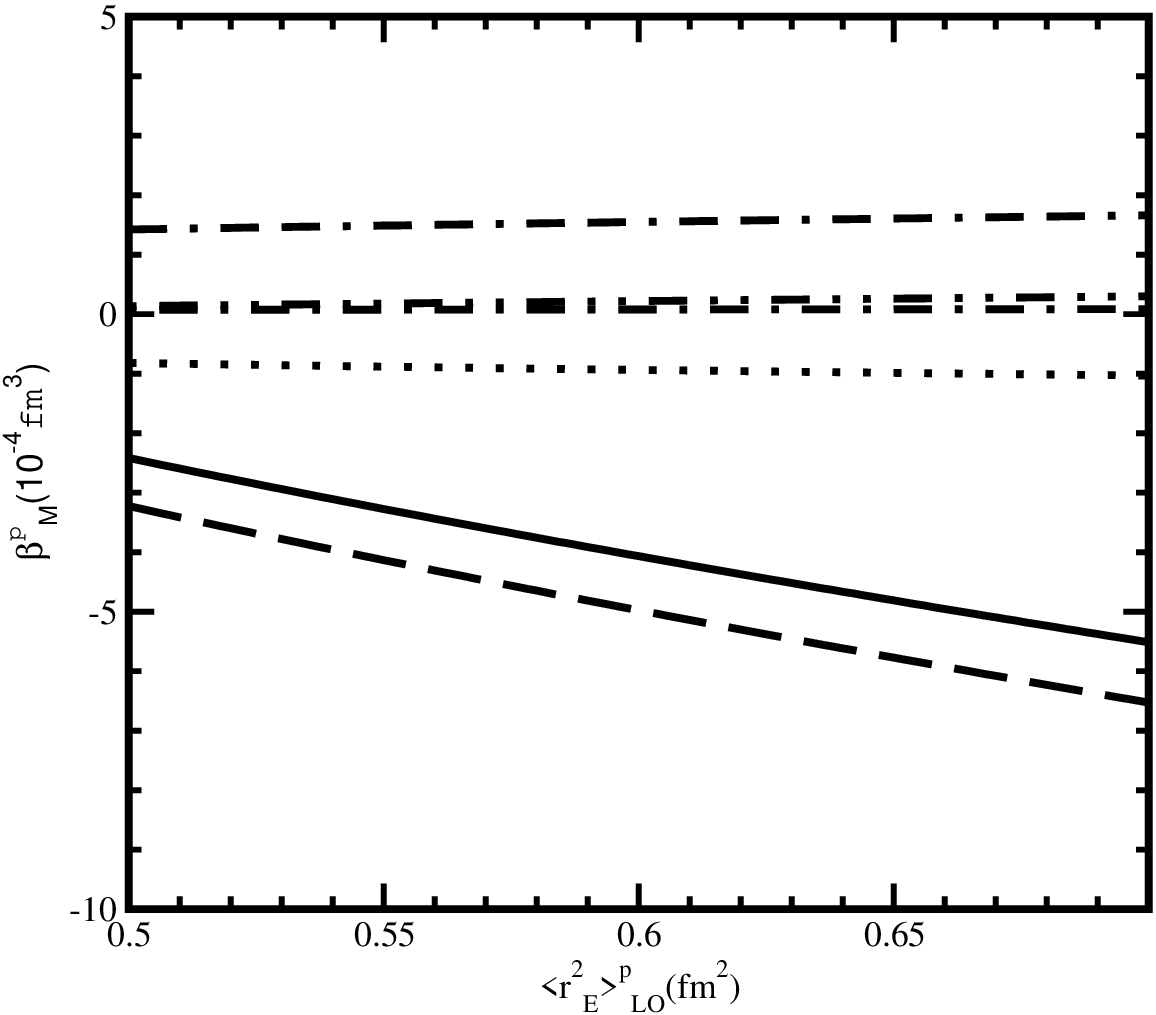,scale=0.40}} 

\vspace*{.45cm}

\centerline{\bf Fig.17}
\end{figure}


\vspace*{3.0cm}

\begin{figure}
\centering{\
\epsfig{figure=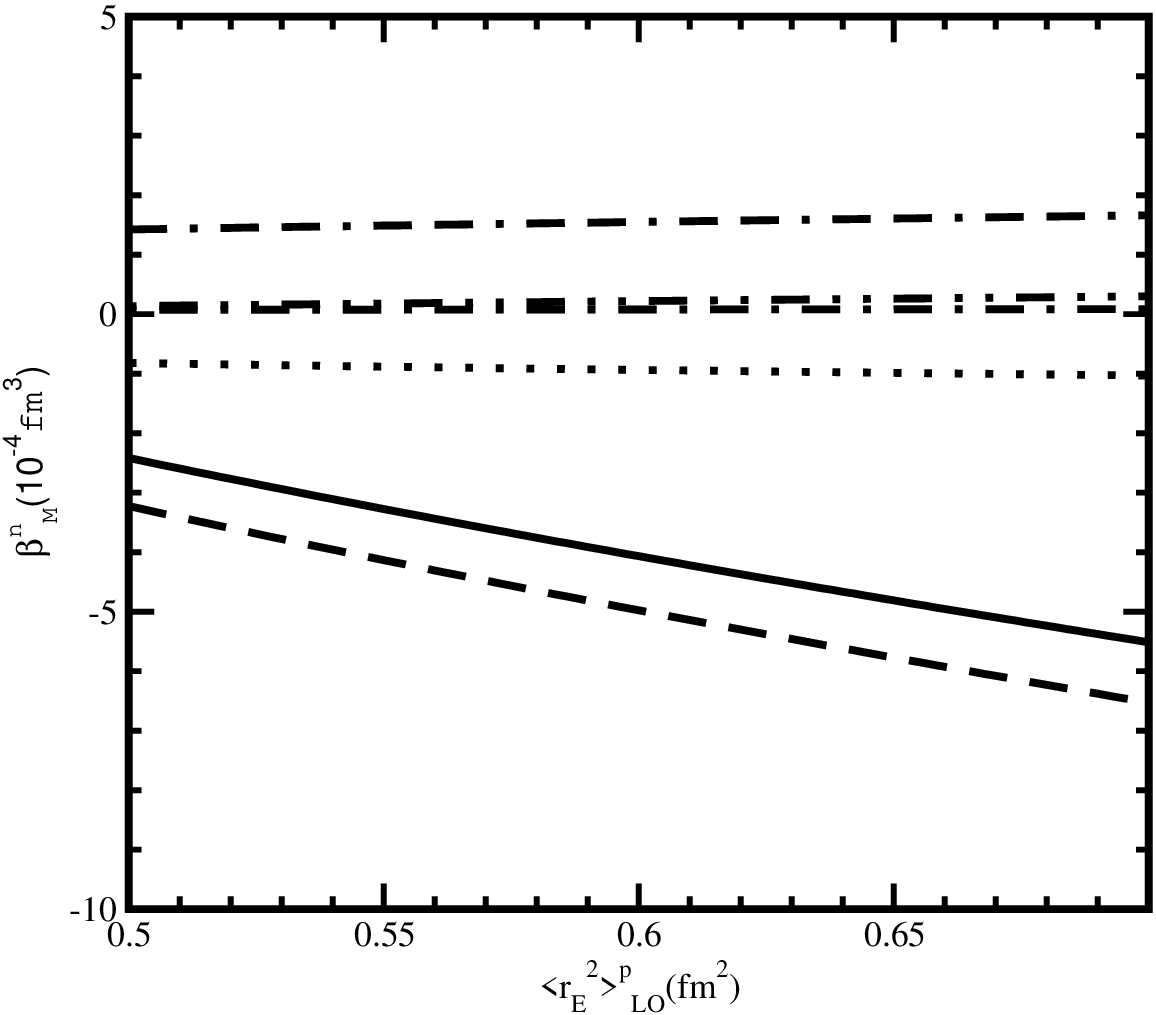,scale=0.40}} 

\vspace*{.45cm}

\centerline{\bf Fig.18}
\end{figure}

\newpage
\begin{table}   
\caption{Results for the nucleon polarizabilities 
(in units of $10^{-4}$fm$^3$).}
\begin{center}
\begin{tabular}{||c|cccc||} 
Approach                &$\alpha_E^p$  &$\beta_M^p$   
                        &$\alpha_E^n$  &$\beta_M^n$ \\ \hline
data~\cite{Schumacher:2005an} &$12.0\pm  0.6$ &$1.9\mp 0.6$   
                          &$12.5\pm 1.7$ &$2.7\mp 1.8$\\ \hline
                          &7.9           &-2.3         
                          &11.0          &-2.0\\ 
ChPT~\cite{Bernard:1991rq}&$10.5\pm 2.0$  &$3.5\pm 3.6$  
                          &$13.6\pm 1.5$ &$7.8\pm 3.6$\\ 
                          &12.6          &1.26         
                          &12.6          &1.26\\ \hline
ChPT + DR~\cite{Lvov:1993ex} &7.3           &-1.8        
                             &9.8           &-0.9 \\ \hline
PCQM                      &10.9         &5.1          
                          &10.9        &1.15\\  
\end{tabular}
\end{center}
\end{table}

\begin{table}   
\caption{Contributions to $\alpha_E^N$ and $\beta_M^N$ from 
the individual diagrams (in units of $10^{-4}$fm$^3$) evaluated 
for the central value $<r^2_E>^P_{\rm LO}= 0.6$ fm$^2$.}
\begin{center}
\begin{tabular}{||c|cccccccc||} 
           &$\alpha_E^p({\rm GS})$ &$\alpha_E^p({\rm ES})$
           &$\alpha_E^n({\rm GS})$ &$\alpha_E^n({\rm ES})$
           &$\beta_E^p({\rm GS})$  &$\beta_M^p({\rm ES})$
           &$\beta_E^n({\rm GS})$  &$\beta_M^n({\rm ES})$ \\ \hline
Fig.1    &0     &0     &0     &0     &0      &11.85   &0      & 7.90 \\ \hline
Fig.2    &2.03  &2.37  &2.03  &2.37  &-0.65  &-4.07   &-0.65  &-4.07 \\ \hline
Fig.3    &4.26  &0     &4.26  &0     &-1.42  &0       &-1.42  &0     \\ \hline
Fig.6    &0     &0     &0     &0     &0.085  &0.02    &0.089   &0.02  \\ \hline
Fig.8    &2.26  &0     &2.26  &0     &-0.71  &0       &-0.71  &0     \\ \hline
Total    &8.55  &2.37  &8.55  &2.37  &-2.70  &7.80    &-2.70  &3.85  \\ 
\end{tabular}
\end{center}
\end{table}

\end{document}